\documentclass[twoside,epsf]{article}
\expandafter\let\csname endequation*\endcsname=\relax 
\usepackage{amsmath,amssymb,amsthm} 
\usepackage{cite}
\usepackage{graphicx}
\usepackage{bm}
\usepackage{upgreek}
\usepackage{braket}
\usepackage{tikz}
\usetikzlibrary{quantikz}
\usepackage{pgfplots}
\usepackage{subcaption}
\pgfplotsset{every tick label/.append style={font=\footnotesize}}
\catcode`\@=11
\long\def\@makefntext#1{
\protect\noindent \hbox to 3.2pt {\hskip-.9pt  
$^{{\eightrm\@thefnmark}}$\hfil}#1\hfill}		

\def\@makefnmark{\hbox to 0pt{$^{\@thefnmark}$\hss}}	
	
\def\ps@myheadings{\let\@mkboth\@gobbletwo
\def\@oddhead{\hbox{}
\rightmark\hfil\eightrm\thepage}   
\def\@oddfoot{}\def\@evenhead{\eightrm\thepage\hfil
\leftmark\hbox{}}\def\@evenfoot{}
\def\sectionmark##1{}\def\subsectionmark##1{}}

\oddsidemargin=\evensidemargin
\newcounter{sectionc}\newcounter{subsectionc}\newcounter{subsubsectionc}
\renewcommand{\section}[1] {\vspace{12pt}\addtocounter{sectionc}{1} 
\setcounter{subsectionc}{0}\setcounter{subsubsectionc}{0}\noindent 
	{\tenbf\thesectionc. #1}\par\vspace{5pt}}
\renewcommand{\subsection}[1] {\vspace{12pt}\addtocounter{subsectionc}{1} 
\setcounter{subsubsectionc}{0}\noindent 
{\bf\thesectionc.\thesubsectionc. {\kern1pt \bfit #1}}\par\vspace{5pt}}
\renewcommand{\subsubsection}[1] {\vspace{12pt}\addtocounter{subsubsectionc}{1}
	\noindent{\tenrm\thesectionc.\thesubsectionc.\thesubsubsectionc.
	{\kern1pt \tenit #1}}\par\vspace{5pt}}
\newcommand{\nonumsection}[1] {\vspace{12pt}\noindent{\tenbf #1}
	\par\vspace{5pt}}

\newcounter{appendixc}
\newcounter{subappendixc}[appendixc]
\newcounter{subsubappendixc}[subappendixc]
\renewcommand{\thesubappendixc}{\Alph{appendixc}.\arabic{subappendixc}}
\renewcommand{\thesubsubappendixc}
	{\Alph{appendixc}.\arabic{subappendixc}.\arabic{subsubappendixc}}

\renewcommand{\appendix}[1] {\vspace{12pt}
        \refstepcounter{appendixc}
        \setcounter{figure}{0}
        \setcounter{table}{0}
        \setcounter{lemma}{0}
        \setcounter{theorem}{0}
        \setcounter{corollary}{0}
        \setcounter{definition}{0}
        \setcounter{equation}{0}
        \renewcommand{\thefigure}{\Alph{appendixc}.\arabic{figure}}
        \renewcommand{\thetable}{\Alph{appendixc}.\arabic{table}}
        \renewcommand{\theappendixc}{\Alph{appendixc}}
        \renewcommand{\thelemma}{\Alph{appendixc}.\arabic{lemma}}
        \renewcommand{\thetheorem}{\Alph{appendixc}.\arabic{theorem}}
        \renewcommand{\thedefinition}{\Alph{appendixc}.\arabic{definition}}
        \renewcommand{\thecorollary}{\Alph{appendixc}.\arabic{corollary}}
        \renewcommand{\theequation}{\Alph{appendixc}.\arabic{equation}}
        \noindent{\tenbf Appendix \theappendixc #1}\par\vspace{5pt}}
\newcommand{\subappendix}[1] {\vspace{12pt}
        \refstepcounter{subappendixc}
        \noindent{\bf Appendix \thesubappendixc. {\kern1pt \bfit #1}}
	\par\vspace{5pt}}
\newcommand{\subsubappendix}[1] {\vspace{12pt}
        \refstepcounter{subsubappendixc}
        \noindent{\rm Appendix \thesubsubappendixc. {\kern1pt \tenit #1}}
	\par\vspace{5pt}}

\newcommand{\textlineskip}{\baselineskip=13pt}
\newcommand{\smalllineskip}{\baselineskip=10pt}

\def\abstracts#1#2#3{{
	\centering{\begin{minipage}{4.5in}\footnotesize\baselineskip=10pt
	\parindent=0pt #1\par 
	\parindent=15pt #2\par
	\parindent=15pt #3
	\end{minipage}}\par}} 

\def\keywords#1{{
	\centering{\begin{minipage}{4.5in}\footnotesize\baselineskip=10pt
	{\footnotesize\it Keywords}\/: #1
	 \end{minipage}}\par}}

\renewenvironment{thebibliography}[1]
{\frenchspacing
	\small\rm\baselineskip=11pt
	\begin{list}{\arabic{enumi}.}
		{\usecounter{enumi}\setlength{\parsep}{0pt}     
			\setlength{\leftmargin}{17pt}  
			\setlength{\rightmargin}{0pt}
			\setlength{\itemsep}{0pt} \settowidth
			{\labelwidth}{#1.}\sloppy}}{\end{list}}

\newcounter{itemlistc}
\newcounter{romanlistc}
\newcounter{alphlistc}
\newcounter{arabiclistc}

\newcommand{\fcaption}[1]{
        \refstepcounter{figure}
        \setbox\@tempboxa = \hbox{\footnotesize Fig.~\thefigure. #1}
        \ifdim \wd\@tempboxa > 5in
           {\begin{center}
        \parbox{5in}{\footnotesize\smalllineskip Fig.~\thefigure. #1}
            \end{center}}
        \else
             {\begin{center}
             {\footnotesize Fig.~\thefigure. #1}
              \end{center}}
        \fi}

\newcommand{\tcaption}[1]{
        \refstepcounter{table}
        \setbox\@tempboxa = \hbox{\footnotesize Table~\thetable. #1}
        \ifdim \wd\@tempboxa > 5in
           {\begin{center}
        \parbox{5in}{\footnotesize\smalllineskip Table~\thetable. #1}
            \end{center}}
        \else
             {\begin{center}
             {\footnotesize Table~\thetable. #1}
              \end{center}}
        \fi}

\def\pmb#1{\setbox0=\hbox{#1}
	\kern-.025em\copy0\kern-\wd0
	\kern.05em\copy0\kern-\wd0
	\kern-.025em\raise.0433em\box0}

\def\fnt#1#2{\footnotetext{\kern-.3em
	{$^{\mbox{\scriptsize #1}}$}{#2}}}

\def\fpage#1{\begingroup
\voffset=.3in
\thispagestyle{empty}\begin{table}[b]\centerline{\footnotesize #1}
	\end{table}\endgroup}

\def\runninghead#1#2{\pagestyle{myheadings}
\markboth{{\protect\footnotesize\it{\quad #1}}\hfill}
{\hfill{\protect\footnotesize\it{#2\quad}}}}
\font\tenrm=cmr10
\font\tenit=cmti10 
\font\tenbf=cmbx10
\font\bfit=cmbxti10 at 10pt

\font\eightrm=cmr8

\def\FigName{figure}%
\newbox\captionbox
\long\def\@makecaption#1#2{%
  \ifx\FigName\@captype
    \vskip\abovecaptionskip
    \setbox\tempbox\hbox{{\figurecaptionfont #1\hskip1em #2}}
	\ifdim\wd\tempbox< 28pc
	\centerline{\box\tempbox}
	\else
	{\figurecaptionfont #1\hskip1em #2\par}
\fi\else
  	\setbox\tempbox\hbox{{\tablecaptionfont #1\hskip1em #2}}
 	\ifdim\wd\tempbox< 28pc 
	\centerline{\box\tempbox}
	\else
	{\tablecaptionfont #1\hskip1em #2\par}%
	\fi   
 \vskip\belowcaptionskip
 \fi}
\InputIfFileExists{psfig.sty}
{\typeout{^^Jpsfig.sty inputed...ok}}{\typeout{^^JWarning: psfig.sty could be be found.^^J}}
\InputIfFileExists{epsfsafe.tex}
{\typeout{^^Jepsfsafe.tex inputed...ok}}
			{\typeout{^^JWarning: epsfsafe.tex could not be found.^^J}}
\InputIfFileExists{epsfig.sty}
{\typeout{^^Jepsfig.sty inputed...ok}}{\typeout{^^JWarning: epsfig.sty could not be found.^^J}}
\InputIfFileExists{epsf.sty}
{\typeout{^^Jepsf.sty inputed...ok}}{\typeout{^^JWarning: epsf.sty could not be found.^^J}}%
\def\fps@figure{tbp}
\def\ftype@figure{1}
\def\ext@figure{lof}
\def\fnum@figure{Fig.\ \thefigure}
\def\qed{\hbox{${\vcenter{\vbox{	          
   \hrule height 0.4pt\hbox{\vrule width 0.4pt height 6pt
   \kern5pt\vrule width 0.4pt}\hrule height 0.4pt}}}$}}

\begin{document}
\setlength{\textheight}{8.0truein}    

\runninghead{Quantum algorithm for the Navier-Stokes equations} 
            {Budinski Lj.}

\normalsize\textlineskip
\thispagestyle{empty}
\setcounter{page}{1}

\vspace*{0.88truein}

\fpage{1}
\centerline{\bf
\begin{tabular}{cc}
& 	QUANTUM ALGORITHM FOR THE NAVIER-STOKES EQUATIONS \\
&   BY USING THE STREAMFUNCTION-VORTICITY FORMULATION \\
&   AND THE LATTICE BOLTZMANN METHOD
\end{tabular}
}
\vspace*{0.37truein}
\centerline{\footnotesize 
LJUBOMIR BUDINSKI}
\vspace*{0.015truein}
\centerline{\footnotesize\it Faculty of Technical Sciences, University of Novi Sad, Trg Dositeja Obradovi\'ca 6}
\baselineskip=10pt
\centerline{\footnotesize\it Novi Sad, 21000,
Serbia\footnote{Associate professor at the Faculty of Technical Sciences, University of Novi Sad, Serbia, ljubab@uns.ac.rs}}
\vspace*{0.225truein}

\vspace*{0.21truein}
\abstracts{
A new algorithm for solving the Navier-Stokes equations (NSE) on a quantum device is presented. For the fluid flow equations the stream function-vorticity formulation is adopted, while the lattice Boltzmann method (LBM) is utilized for solving the corresponding system of equations numerically for one time step. Following the nature of the lattice Boltzmann method, the proposed quantum algorithm consists of five major sections: initialization, collision, propagation, boundary condition implementation, and calculation of macroscopic quantities. The collision and boundary condition step is quantumly implemented by applying the standard-form encoding approach, while the quantum walk procedure is applied for the propagation step. The algorithm is implemented by using IBM's quantum computing software development framework Qiskit, while for the verification purposes two-dimensional (2D) cavity flow is simulated and compared with classical code.}{}{}

\vspace*{10pt}
\vspace*{10pt}
\vspace*{1pt}\textlineskip	
\vspace*{-0.5pt}
\noindent
\section{Introduction}
Solving differential equations in high-dimensional spaces presents one of the major obstacles in the scientific and engineering community. The need to accurately and efficiently solve a variety of physical processes in large domains described by the complex differential equations has driven scientists and engineers towards more computationally efficient architectures. In that context the development of parallel computing processors and platforms, like Xeon Phi processors (Intel) or CUDA-enabled graphics processing unit (Nvidia), significantly increased efficiency of the mathematical models in the past decade. The next computational architecture which can offer an extensive speedup in comparison with parallel-based processors are computers harvesting the principles of the quantum mechanics - quantum computers. Exploiting the basic properties of the quantum world, like superposition and entanglement, these computers can deliver exponential speedup in various fields of science like machine learning \cite{Schuld1,garg,Sharma,Schuld2,Florian,Wittek}, chemistry \cite{Cao,Egger}, finance \cite{Bouland} and linear algebra \cite{Harrow,Ambainis,Qian}.

Due to the extensive potential speedup provided by the quantum technology, it is legit to ask if the same technology could be used to solve problems involving processes described by the complex partial differential equations (PDE), like weather forecasting, fluid flow, turbulence, etc. The first attempt to solve a linear system of equations on quantum computer is done by Harrow et al.\ \cite{Harrow}, where the exponential speedup over the best known classical algorithms is achieved by the algorithm that runs in poly($\log N$). The improvement of the original Harrow-Hassidim-Lloyd algorithm (HHL) in terms of calculation time from $ \mathcal{O}(\kappa \log N) $ to $ \mathcal{O}(\kappa \log^3 \kappa \log N) $, where $ N $ is the size
of the linear system and $ \kappa $ is the condition number of the system of equations, is presented by Ambainis \cite{Ambainis}, while solutions for the state preparation, solution readout and the condition number, as features that limit promised exponential speedup, is provided by Clader \cite{Clader}. Additional exponential improvement of the dependence on the precision parameter in the phase estimation part of the HHL algorithm is introduced by Childs et al.\ \cite{Childs}. To solve the systems of nonlinear
algebraic equations, authors Peng et al.\ \cite{Qian} proposed a quantum algorithm based on optimization strategy, where the approximate solutions to the equations were found by searching the entire space of computation basis. To extend the applicability of quantum devices to the domain of differential equations, Berry \cite{Berry} proposed a quantum algorithm for solving the inhomogeneous sparse linear differential
equations by encoding the differential equations as a linear system by discretization, and then solving the corresponding algebraic system of equations using the HHL algorithm. To improve precision dependence from poly($1/\varepsilon$) to poly($\log(1/\varepsilon)$), where $ \varepsilon $ represents the solution error, Berry et al.\ \cite{Berry2} applied the linear combinations of unitaries approach to solving the linear system that encodes a truncated Taylor series. Examples of quantumly solved differential equations describing some physical processes by using the presented quantum algorithms can be found in \citen{Cao2,Tao,Dutta,Costa,Doronin}. Besides the grid-based approach, where discretization of the differential equations (finite difference method - FDM) to obtain a system of linear equations is performed, Childs and Liu \cite{Childs3} proposed a quantum algorithm based on linear combinations of basis functions to approximate the solution globally (spectral method), while high-precision quantum algorithm for linear partial differential equations with the complexity poly($\log(1/\varepsilon),d $) is done by Childs et al.\ \cite{Childs2}. Alternative approaches for solving the differential equations on Noisy Intermediate-
Scale Quantum devices (NISQ), like the variational algorithm, where the quantum computer is used to prepare the solution vector using a shallow sequence of parameterized quantum gates while the measurements are performed on the solution to evaluate the quality defined in terms of a loss function is also extensively investigated \cite{Lee,Huang,Lubasch,bravoprieto,liu}. Furthermore, the first application of the continuous-variable (CV) quantum computer on solving one-dimensional ordinary differential equation is done by Knudsen and Mendl \cite{Knudsen}.

In this work quantum algorithm for the two-dimensional (2D) Navier-Stokes equation is proposed. To solve this system of non-linear partial differential equations (the continuity equation and two momentum equations) on quantum devices, a suitable temporal and spatial discretization model is required. Theoretically, one of the possible routs to follow, in general, is to follow the previously proposed procedures by Barry \cite{Berry}, where the NSE could be firstly converted into ordinary differential equations (ODE) by using appropriate discretization procedure, and then solved by applying some of the corresponding quantum algorithms. This procedure was utilized by the Gaitan \cite{Gaitan} in order to solve the one-dimensional NSE for flow through a convergent-divergent nozzle, where the quantum amplitude estimation algorithm (QAEA) \cite{Brassard} is used to obtain the approximate ODE solution. Application of the variational approach for solving solving non-linear differential equations with differentiable quantum circuits, where for the test example quasi-1D Navier-Stokes equation are used, is done by Kyriienko et. al \cite{Kyriienko}, while the hybrid classical/quantum algorithm based on a Vortex-In-Cell formulation of the NSE is performed by the Steijl and Barakos \cite{Steijl}. Another numerical approach, which can relax the non-linearity of the NSE without jeopardizing the stability, accuracy, and efficiency of the final solution on the one hand, and provide simple mathematical structure suitable for parallel computation, and therefore allow quantum computing, on the other hand, is the lattice Boltzmann method (LBM) \cite{Rivet,Rothman,Chen,Mohamad}. The development of the first quantum simulator suitable for solving fluid dynamic transport phenomena based on a lattice kinetic formalism is done by Mezzacapo et al. \cite{Mezzacapo}, while an attempt to solve the simpler form of the lattice Boltzmann equation on a quantum computer is done by Todorova and Steijl \cite{Blaga}. Only the propagation step using the quantum walk procedure was applied in this work, which in turn reduced the applicability of the LBM significantly. However, due to the linear nature of the procedure, it can run all required time steps in simulation on a quantum processor without repeated exchange with classical computer each times-step. Application of the quantum walk on two-dimensional lattice, where the multi particle interaction is modeled by using the collision procedure from the lattice gas automaton HPP model (Hardy, Pomeau and de Pazzis) is done by the Costa et al. \cite{Costa}, while some even simpler flows by using the lattice-gas
model are quantumly modeled by Micci and Yepez \cite{Micci}. 

The extension of previous work \cite{Budinski}, where the quantum algorithm for advection-diffuison equation in framework of the LBM is given, on solving the NSE equations is presented in this paper. To solve unsteady,  incompressible two-dimensional NSE on a quantum computer by utilizing the lattice Boltzmann method, where the extension to the three-dimensional case is straightforward, the non-linearity of the  basic form of equilibrium distribution function \cite{Rivet} is significantly reduced by using the stream function-vorticity formulation in this work. Two partial differential equations are numerically modeled by using the simpler LBM configuration and quantumly solved on IBM's open-source quantum computing software development framework Qiskit \cite{Qiskit}. Due to the presence of the non-linear term in the equilibrium distribution function, the presented algorithm is restricted to one time step, after which measurements and re-initialization of the state for the purpose of new time step is required. Implementation on Qiskit is done by using the standard gate set and built-in procedures and algorithms, while for the test case a two-dimensional (2D) cavity flow is simulated and compared with the classical code. The solution of the unsteady, incompressible high-dimension NSE on a quantum computer by using the complete propagation-collision form of the LBM is the major contribution of this work, which is to the best of the author's knowledge the very first attempt to solve the system of 2D(3D) Navier-Stokes equations on the quantum device.

\section{The mathematical model}

\subsection{Stream function-Vorticity Formulation} 

For unsteady, incompressible two-dimensional flows with constant fluid properties, the Navier–Stokes equations can be simplified by introducing the vorticity $ \omega $ and stream function $ \psi $ as dependent variables. This type of flow can be described by the governing vorticity transport and Poisson equations,

\begin{equation}
	\frac{\partial \omega}{\partial t}+u\frac{\partial \omega}{\partial x}+v\frac{\partial \omega}{\partial y}=\nu \left(\frac{\partial^2 \omega}{\partial x^2}+\frac{\partial^2 \omega}{\partial y^2}\right),
	\label{1}
\end{equation} 

\begin{equation}
	\frac{\partial^2 \psi}{\partial x^2}+\frac{\partial^2 \psi}{\partial y^2}=-\omega.
	\label{2}
\end{equation}    
where $ t $ is time, $ x,y $ are the Cartesian coordinates, $ u $ and $ v $ are the flow velocity components in the $ x$ and $y $ direction, respectively, while $ \nu $ is the kinematic viscosity. Furthermore, the stream function is given by

\begin{equation}
	\frac{\partial \psi}{\partial y}=u, \frac{\partial \psi}{\partial x}=-v,
	\label{3}
\end{equation}                     
which define lines with constant $ \psi $, that are everywhere parallel to the flow. 

\subsection{The lattice Boltzmann model}

The evolution of particle distribution function can be described by the single relaxation time lattice Boltzmann equation \cite{Rivet,Rothman} as,

\begin{equation}
	f_\alpha(\mathbf{x}+\textbf{e}_\alpha\Delta t,t+\Delta t)=\left( 1-\varepsilon\right)f_\alpha(\mathbf{x},t) +\varepsilon f^{eq} _\alpha+\Delta t w_{\alpha} S,
	\label{4}
\end{equation}
where $ \mathbf{x}=\left( x,y\right) $ is the position vector defined by Cartesian coordinates in the two-dimensional space, $f_\alpha$ is the particle distribution function along the $\alpha$ link,  $f^{eq} _\alpha$ is the local equilibrium distribution function, $ \mathbf{e}_\alpha$ is the particle velocity vector, $ S $ is the force (source) term, $ w_{\alpha} $ is the weighting factor in the direction $ \alpha $,  $t$ is time, $\Delta t$ is the time step and $ \varepsilon $=$ \Delta t $/$\tau$, where $\tau$ is the single relaxation time. 

\begin{figure}
	\centering
	\includegraphics[width=8.0cm]{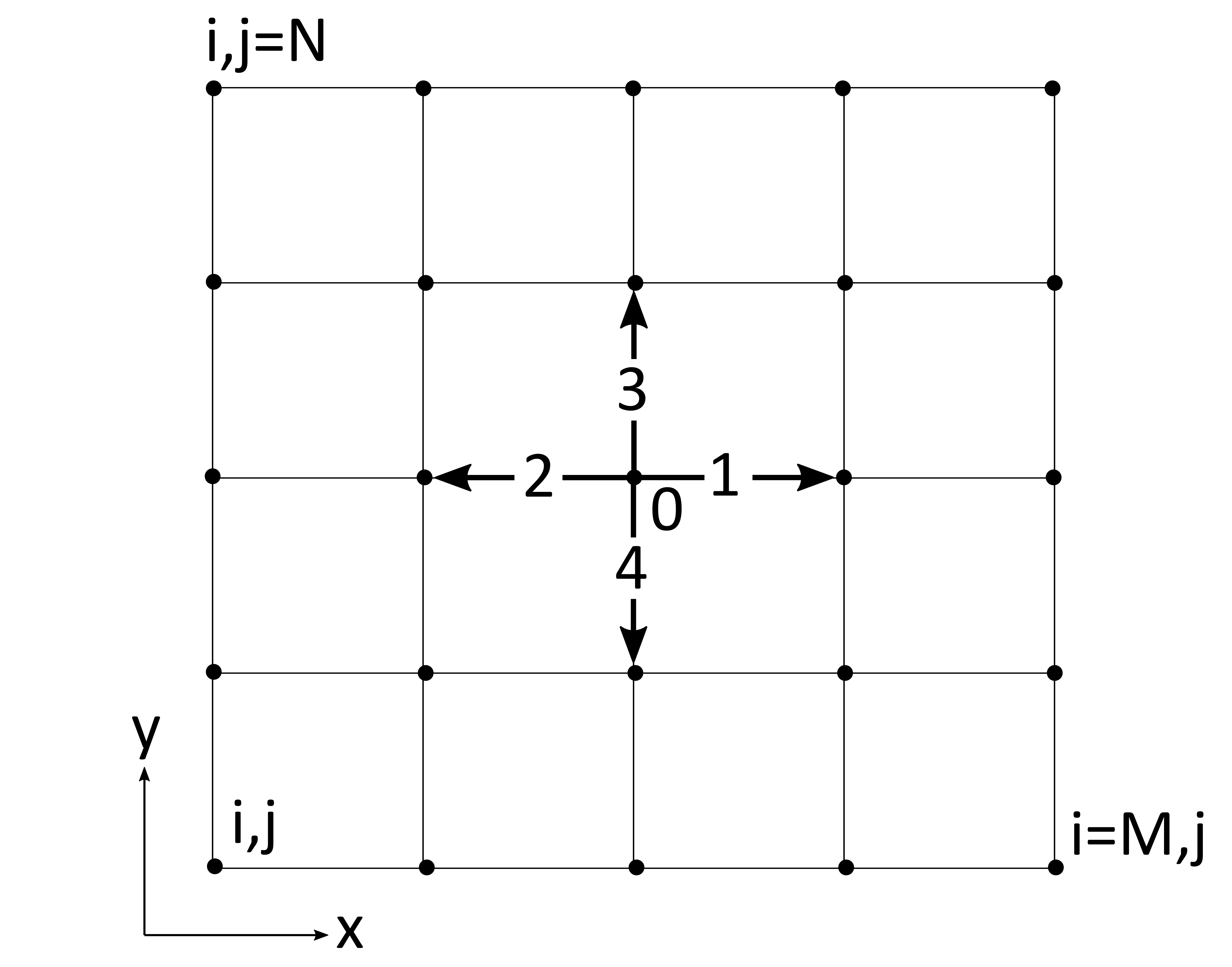}
	\fcaption{D2Q5 lattice configuration.}
	\label{fig:FIG-1}
\end{figure}

To solve Eqs.(\ref{1})-(\ref{2}) in the two-dimensional plane, a D2Q5 configuration of the lattice Boltzmann method with the rest particle is used (Fig.\ref{fig:FIG-1}). The velocity vectors for this LBM scheme are defined as,

\begin{equation}
	\displaystyle e_\alpha=\left\{\begin{array}{ll}\displaystyle (0,0),&\alpha=0\\
		\\
		\displaystyle \left(\pm e_x,0\right),&\alpha=1,2\\
		\\
		\displaystyle \left(0,\pm e_y \right),&\alpha=3,4
	\end{array}\right.
	\label{5}
\end{equation} 
where $ e_x=e_y=\Delta x(y)/\Delta t $. Since the equilibrium distribution function $ f_{\alpha}^{eq} $ defines the final form of the macroscopic equation being modeled, in case of the vorticity transport equation (Eq.(\ref{1})), the equilibrium distribution function is formulated as
\begin{equation}
	f_{\alpha}^{eq}\left(\mathbf{x},t\right)=w_\alpha\omega\left(\mathbf{x},t\right)\left(1+\frac{e_\alpha\cdot\overrightarrow{u}}{c_s^2}\right),
	\label{6}
\end{equation}
while in the case of the Poisson equation (Eq.(\ref{2})), where only the diffusion part is present, the equilibrium function takes a much simpler form 
\begin{equation}
	g_{\alpha}^{eq}\left(\mathbf{x},t\right)=w_\alpha\psi\left(\mathbf{x},t\right).
	\label{7}
\end{equation}
The corresponding weight coefficients and the speed of sound are set to $ w_0=2/6 $, $ w_{1,2,3,4}=1/6 $ and  $ c_s=1/\sqrt{3} $, respectively, while the advection  velocity vector for the 2D case is $ \overrightarrow{u}=ui+vj $, where $ i $ and $ j $ are unit vectors along the $ x $ and $ y $ direction. 

One of the major advantages offered by the lattice Boltzmann method lies in simple and efficient implementation of Eq.(\ref{4}). In general, the evolution process can be decomposed into two basic steps, collision and streaming. Hence, for the vorticity transport equation the collision step is calculated as
\begin{equation}
	\hat{f}_\alpha(\mathbf{x},t)=\left( 1-\varepsilon\right)f_\alpha(\mathbf{x},t) +\varepsilon f^{eq} _\alpha,
	\label{8}
\end{equation}
while for the Poisson equation, it can be formulated as 
\begin{equation}
	\hat{g}_\alpha(\mathbf{x},t)=\left( 1-\varepsilon\right)g_\alpha(\mathbf{x},t) +\varepsilon g^{eq} _\alpha+\Delta t w_{\alpha} S,
	\label{9}
\end{equation}
where the source term is calculated as $ S=-\omega $. In the streaming step propagation of the relaxed distribution functions $ f_\alpha $ and $ g_\alpha $ along the links $ \alpha $ is conducted as
\begin{equation}
	{f}_\alpha,g_\alpha(\mathbf{x}+\textbf{e}_\alpha\Delta t,t+\Delta t)=\hat{f}_\alpha,\hat{g}_\alpha(\mathbf{x},t).
	\label{10}
\end{equation}
At the end of each time step the macroscopic variable for vorticity $ \omega\left(\mathbf{x},t\right) $ and the stream function $ \psi\left(\mathbf{x},t\right) $ is calculated as
\begin{equation}
	\omega\left(\mathbf{x},t\right)=\sum_{\alpha} f_\alpha\left(\mathbf{x},t\right), \psi\left(\mathbf{x},t\right)=\sum_{\alpha} g_\alpha\left(\mathbf{x},t\right).
	\label{11}
\end{equation}
In order to further simplify the lattice Boltzmann procedure, in this work parameter $ \varepsilon $ is,  according to \cite{Zhou}, in both cases set to $ 1.0 $. As a result the collision step (Eq.(\ref{8}) and Eq.(\ref{9})) includes only equilibrium distribution function, which significantly reduces complexity of the entire procedure. The detailed recovery of Eqs.(\ref{1})-(\ref{2}) by applying the Chapman-Enskog expansion procedure can be found in \cite{Yan}. 

\subsection{Boundary conditions}

The boundary conditions on solid boundaries can be derived from the constant stream function assumption $ \psi_{bound}=0 $, and the Taylor expansion of the vorticity equation (Eq.(\ref{2})). For example, in the case of the top wall (Fig.\ref{fig:FIG-1}), where Eq.(\ref{2}) is according to the constant stream function condition reduced to 
\begin{equation}
	\left(\frac{\partial^2 \psi}{\partial y^2}\right)_{i,N}=-\omega_{i,N},
	\label{12}
\end{equation}
the boundary condition in terms of the unknown stream function $ \psi $ can be defined as
\begin{equation}
	\psi_{i,N-1}=\psi_{i,N}-\left(\frac{\partial \psi}{\partial y}\right)_{i,N}\Delta y+\left(\frac{\partial^2 \psi}{\partial y^2}\right)_{i,N}\frac{\Delta y^2}{2}+....\ .
	\label{13}
\end{equation}     
By inserting Eq.(\ref{3}) and Eq.(\ref{12}) into Eq.(\ref{13}), final form of the boundary equation in terms of unknown vorticity is defined as
\begin{equation}
	\omega_{i,N}=2\frac{-\psi_{i,N-1}}{\Delta y^2}-\frac{2U}{\Delta y},
	\label{14}
\end{equation}
where $ U $ is the prescribed velocity on the solid boundary. 

To implement these boundary conditions into the lattice Boltzmann procedure, an appropriate formulation in terms of distribution function $ f_{\alpha} $ is required. In the case of the D2Q5 lattice configuration there is just one unknown distribution function per boundary, which can be defined as distribution function along the link $\alpha$ that is directed inwards. For example,  the unknown distribution functions on the left and right boundary are $ f_{1} $ and $ f_{2} $, respectively, while in the case of top and bottom boundary distribution functions $ f_{4} $ and $ f_{3} $ should be determined, respectively. Hence, the unknown distribution function on the top wall $ f_4 $ can be calculated by combining Eq.(\ref{11}) and Eq.(\ref{14}) as
\begin{equation}
	f_{4(i,N)}=-\left(f_{0(i,N)}+f_{1(i,N)}+f_{2(i,N)}+f_{3(i,N)}\right)-2\frac{-\psi_{i,N-1}}{\Delta y^2}-\frac{2U}{\Delta y}.
	\label{15}
\end{equation}  
A similar procedure can be applied for the calculation of the unknown distribution functions $ g_{\alpha} $ as well, where by combining Eq.(\ref{11}) and the zero stream function assumption $ \psi_{bound}=0 $, a much simpler form of the final equation is obtained,
\begin{equation}
	g_{4(i,N)}=-\left(g_{0(i,N)}+g_{1(i,N)}+g_{2(i,N)}+g_{3(i,N)}\right).
	\label{16}
\end{equation}     

\section{The quantum algorithm}
To develop and test the quantum algorithm for solving the stream function-vorticity formulation of the NSE by using the LBM, the IBM's quantum computing software development framework Qiskit \cite{Qiskit} is used in this work. In general, the methodology for establishing the quantum algorithm is primarily determined by the structure of the lattice Boltzmann method, and can be defined through five major steps: initialization, collision, propagation, boundary condition implementation and, calculation of macroscopic quantities (Fig.\ref{fig:FIG-2}). For each of these steps a corresponding quantum algorithm has been developed and implemented on Qiskit's statevector simulator by using its gate set and built-in procedures, while for to the practical limitation in terms of available qubits, the resulting components of velocities are calculated on a classical computer by using Eq.(\ref{3}). However, some practical guidelines on solving Eq.(\ref{3}) on the quantum device are consequently provided. After one time step simulation is finished, the measurements and re-initialization of the quantum state serving as the input for next time step is required. This condition introduces additional restriction, where, due to the presence of non-linear term in equilibrium function, performing the state tomography after each time step is obligatory.
\begin{figure}
	\centering
	\begin{quantikz}
		\lstick[wires=3]{$\ket{0}^{\otimes n}_q$} & \gate[style={fill=red!40},label style=black,5, disable auto height][1.0cm]{\arraycolsep=1.4pt\def\arraystretch{0.6} \begin{array}{c} \rotatebox{90}{G} \\ \rotatebox{90}{N} \\ \rotatebox{90}{I} \\ \rotatebox{90}{D} \\ \rotatebox{90}{O} \\ \rotatebox{90}{C} \\ \rotatebox{90}{N} \\ \rotatebox{90}{E} \end{array}} \qwbundle [alternate]{} \slice{$ \ket{\phi_{0}} $} &  \qwbundle [alternate]{}  & \gate[style={fill=blue!40},label style=black,5, disable auto height][1.0cm]{\arraycolsep=1.4pt\def\arraystretch{0.6} \begin{array}{c} \rotatebox{90}{N} \\ \rotatebox{90}{O} \\ \rotatebox{90}{I} \\ \rotatebox{90}{S} \\ \rotatebox{90}{I} \\ \rotatebox{90}{L} \\ \rotatebox{90}{L} \\ \rotatebox{90}{O} \\ \rotatebox{90}{C} \end{array}} \qwbundle [alternate]{} & \qwbundle [alternate]{} \slice{$ \ket{\phi_{1}} $} & \qwbundle [alternate]{}  & \gate[style={fill=green!40},label style=black,5, disable auto height][1.0cm]{\arraycolsep=1.4pt\def\arraystretch{0.6} \begin{array}{c} \rotatebox{90}{N} \\ \rotatebox{90}{O} \\ \rotatebox{90}{I} \\ \rotatebox{90}{T} \\ \rotatebox{90}{A} \\ \rotatebox{90}{G} \\ \rotatebox{90}{A} \\ \rotatebox{90}{P} \\ \rotatebox{90}{O} \\ \rotatebox{90}{R} \\ \rotatebox{90}{P} \end{array}} \qwbundle [alternate]{} & \qwbundle [alternate]{} \slice{$ \ket{\phi_{2}} $} & \qwbundle [alternate]{}  & \gate[style={fill=yellow!40},label style=black,5, disable auto height][1.0cm]{\arraycolsep=1.4pt\def\arraystretch{0.6} \begin{array}{c} \rotatebox{90}{S} \\ \rotatebox{90}{O} \\ \rotatebox{90}{R} \\ \rotatebox{90}{C} \\ \rotatebox{90}{A} \\ \rotatebox{90}{M} \end{array}} \qwbundle [alternate]{} & \qwbundle [alternate]{} \slice{$ \ket{\phi_{3}} $} & \qwbundle [alternate]{}  & \gate[style={fill=orange!40},label style=black,5, disable auto height][1.0cm]{\arraycolsep=1.4pt\def\arraystretch{0.6} \begin{array}{c} \rotatebox{90}{Y} \\ \rotatebox{90}{R} \\ \rotatebox{90}{A} \\ \rotatebox{90}{D} \\ \rotatebox{90}{N} \\ \rotatebox{90}{U} \\ \rotatebox{90}{O} \\ \rotatebox{90}{B} \end{array}} \qwbundle [alternate]{} & \qwbundle [alternate]{} \slice{$ \ket{\phi_{4}} $} & \qwbundle [alternate]{}  \\ 
		& \qw &  \qw  & \qw & \qw & \qw  & \qw &\qw  & \qw & \qw  & \qw & \qw & \qw & \qw & \qw \\
		& \qw &  \qw  & \qw & \qw & \qw  & \qw &\qw  & \qw & \qw  & \qw & \qw & \qw & \qw & \qw \\ [1.5cm]
		\lstick{$\ket{0}_{a1}$} & \qw &  \qw  & \qw & \qw & \qw  & \qw &\qw  & \qw & \qw  & \qw & \qw & \qw & \qw & \qw \\
		\lstick{$\ket{0}_{a2}$} & \qw & \qw & \qw \qw & \qw & \qw & \qw & \qw & \qw & \qw & \qw  & \qw & \qw & \qw & \qw
	\end{quantikz}
	\fcaption{Quantum circuit for solving the 2D Navier-Stokes equations by using the D2Q5 lattice Boltzmann model. For the typesetting quantum circuit diagrams, the Quantikz package \cite{Alastair} is used.}
	\label{fig:FIG-2}
\end{figure}       
\subsection{The quantum-circuit construction}
\emph{Encoding} - In the first step initial values of vorticity $ \omega\left(\mathbf{x},0\right) $, stream function $ \psi\left(\mathbf{x},0\right) $, source term $ S\left(\mathbf{x},0\right)  $ (Eq.(\ref{9})), and boundary condition related to Eq.(\ref{14}), defined in vector form as 

\begin{equation}
	\bm{\uplambda}=\left[\begin{array}{ll}\displaystyle \left[(\omega_{0,i,j},\dots,\omega_{0,N-1,M-1}), \dots ,(\omega_{4,i,j},\dots,\omega_{4,N-1,M-1})\right]^T,\\
		\displaystyle \left[(\psi_{0,i,j},\dots,\psi_{0,N-1,M-1}), \dots ,(\psi_{4,i,j},\dots,\psi_{4,N-1,M-1})\right]^T,\\
		\displaystyle \left[(S_{0,i,j},\dots,S_{0,N-1,M-1}), \dots ,(S_{4,i,j},\dots,S_{4,N-1,M-1})\right]^T,\\
		\displaystyle \left[(\omega_{b,i,j},\dots,\omega_{b,N-1,M-1})\right]^T,\\
		
		\left[(\omega_{0,i,j},\dots,\omega_{0,N-1,M-1}), \dots ,(\omega_{4,i,j},\dots,\omega_{4,N-1,M-1})\right]^T,\\
		\displaystyle \left[(\psi_{0,i,j},\dots,\psi_{0,N-1,M-1}), \dots ,(\psi_{4,i,j},\dots,\psi_{4,N-1,M-1})\right]^T,\\
		\displaystyle \left[(S_{0,i,j},\dots,S_{0,N-1,M-1}), \dots ,(S_{4,i,j},\dots,S_{4,N-1,M-1})\right]^T,\\
		\displaystyle \left[(\omega_{b,i,j},\dots,\omega_{b,N-1,M-1})\right]^T
	\end{array}\right],  
	\label{17}
\end{equation}
is encoded into the quantum state as 
\begin{equation}
	\ket{\phi_{0}}=\ket{0}_{a1} \ket{0}_{a2} \sum_{k=0}^{32(N\times M)-1} \lambda_{\alpha,k}/\| \lambda \| \ket{k}_q.
	\label{18}
\end{equation}
To encode this initial vector into the quantum state a quantum register $ q $ with $ \log_2(32(N\times M)) $ qubits is required, while the two ancillary registers, denoted with the indexes $ a_1 $ and $ a_2 $, are introduced for calculation purposes only. This initialization of an arbitrary vector into the corresponding quantum state is achieved by using the reverse iterative procedure proposed by Shende et al. \cite{Shende} and it is part of the Qiskit \cite{Qiskit} framework. It should be noted that during the simulation this type of encoding is performed before each time step, i.e. after the re-initialization of the quantum state at the end of the time step is conducted.    

\emph{Collision} - In the collision step, where by using $ \varepsilon=1.0 $, Eq.(\ref{8}) and Eq.(\ref{9}) are reduced to a much simpler form, i.e. $ \hat{f}_\alpha(\mathbf{x},t)=f^{eq} _\alpha $ and $ \hat{g}_\alpha(\mathbf{x},t)=g^{eq} _\alpha+\Delta t w_{\alpha} S $, respectively, equilibrium functions $ f^{eq} _\alpha $ and $ g^{eq} _\alpha $ are first calculated by performing multiplication of vector $ \bm{\uplambda} $ with the diagonal matrix $ A $ having entries corresponding to Eq.(\ref{6}) and Eq.(\ref{7}). In that sense matrix $ A $ is constructed of 32 $ M \times M $ diagonal blocks, in which entries of the first 5 diagonal blocks correspond to vorticity equilibrium function $ f_{\alpha} $ (Eq.\ref{6}) without the vorticity term $ \omega\left(\mathbf{x},t\right) $, the second group of 5 diagonal blocks refers to the stream function equilibrium function $ g_{\alpha} $ (Eq.\ref{7}) excluding the stream term $ \psi\left(\mathbf{x},t\right) $, while the third group of 5 diagonal blocks and the fourth diagonal $ M \times M $ block correspond to the source term $ S $, and the boundary condition for the vorticity, respectively. For the source term $ S $ only weight coefficient $ w_{\alpha} $ is used, while in case of the boundary condition prefactor $ 1/\sqrt{2} $ is applied. According to Eq.(\ref{17}) the second half of matrix $ A $ is populated by simply copying the entries from the first half of the matrix. The upper and lower half of the matrix are denoted as $ A_1 $ and $ A_2 $, respectively. To implement this non-unitary matrix into the quantum algorithm the linear combination of unitaries approach \cite{Low} is applied. To perform this task unitary operators $ B_1 $ and $ B_2 $ are constructed from matrix $ A $, using the approach from \cite{Tao} first, where $ B_1=A_1+i\sqrt{I-A_1^2} $ and $ B_2=A_2-i\sqrt{I-A_2^2} $ are the upper and lower half of operator $ B $, respectively. The collision step is then implemented by applying the operator in the form
\begin{equation}
	(I_{a2} \otimes H_{a1} \otimes I_q)(I_{a2} \otimes SWAP_{(a1-qn)} \otimes I_{q\-1})(I_{a2} \otimes H_{a1} \otimes B_q),
	\label{19}
\end{equation} 
where $ H $ is the Hadamard operator, $ SWAP $ is the operator for swapping the states between two qubits, while $ I $ denotes the identity operator. For the implementation of operator $ B $ a Qiskit's diagonal subroutine based on \cite{Shende} is used, while the corresponding quantum circuit for the collision step is given in Fig.(\ref{fig:FIG-3}). By applying the operator from Eq.(\ref{19}), a state $ \phi_0 $ evolved into:
\begin{equation}
	{\begin{array}{rl}\ket{\phi_{1}^*}&=\ket{0}_{a2} \ket{0}_{a1}\left(\displaystyle \sum_{k=0}^{5(N\times M)-1} a_{k,k} \lambda_{\alpha,k}/\| \lambda \| \ket{k}_q+\sum_{k=5(N\times M)-1}^{10(N\times M)-1} b_{k,k} \lambda_{\alpha,k}/\| \lambda \| \ket{k}_q\right)\\
			&+\ket{0}_{a2} \ket{0}_{a1}\left(\displaystyle \sum_{k=10(N\times M)-1}^{15(N\times M)-1} c_{k,k} \lambda_{\alpha,k}/\| \lambda \| \ket{k}_q+\sum_{k=15(N\times M)-1}^{16(N\times M)-1} d_{k,k} \lambda_{k}/\| \lambda \| \ket{k}_q\right)\\
			&+\ket{0}_{a2} \ket{0}_{a1}\left(\displaystyle \sum_{k=16(N\times M)-1}^{21(N\times M)-1} a_{k,k} \lambda_{\alpha,k}/\| \lambda \| \ket{k}_q+\sum_{k=21(N\times M)-1}^{26(N\times M)-1} b_{k,k} \lambda_{\alpha,k}/\| \lambda \| \ket{k}_q\right)\\
			&+\ket{0}_{a2} \ket{0}_{a1}\left(\displaystyle \sum_{k=26(N\times M)-1}^{31(N\times M)-1} c_{k,k} \lambda_{\alpha,k}/\| \lambda \| \ket{k}_q+\sum_{31(N\times M)-1}^{32(N\times M)-1} d_{k,k} \lambda_{k}/\| \lambda \| \ket{k}_q\right)\\
			&+\ket{0}_{a2} \ket{1_\lambda^\bot}_{a1q},
	\end{array}}
	\label{20}
\end{equation}
where $ a_{k,k} $ denotes entries of the diagonal matrix $ A $ that correspond to the vorticity equilibrium distribution function $ f_{\alpha}^{eq} $ (Eq.\ref{6}), $ b_{k,k} $ corresponds to the stream function equilibrium distribution function $ g_{\alpha}^{eq} $ (Eq.\ref{7}), $ c_{k,k} $ denotes the weight coefficients $ w_{\alpha} $ multiplying the source term $ S $ in Eq.(\ref{9}), while in the case of $ d_{k,k} $  prefactor $ 1/\sqrt{2} $ is used. In general, two copies of the $ f_{\alpha}^{eq} $, $ g_{\alpha}^{eq} $, source term $ S $ and boundary condition are encoded into the quantum state $ \phi_1^* $, where $ \ket{1_\lambda^\bot}_{a1q} $ denotes some orthogonal state of lesser interest.

According to Eq.(\ref{9}), the second part of the collision step corresponds to the point-wise addition of the source term $ \Delta t w_{\alpha} S $ (in this paper $ \Delta t=1.0 $) with the stream function equilibrium distribution function $ g_{\alpha}^{eq} $ calculated previously. This part of the collision step includes the identification and shifting of each part of state $ \phi_1^* $ which refers to the source term against the corresponding equilibrium distribution function $ g_{\alpha}^{eq} $ in the first step, and then application of the Hadamard operator in the second step. Since for each link $ \alpha $ a $ N \times M $ dimensional quantum sub-state corresponding to the equilibrium distribution function $ g_{\alpha}^{eq} $ is constructed (Eq.(\ref{20})), there are 5 separate identification-shifting procedures in total, required to 'align' each source term with the corresponding $ g_{\alpha}^{eq} $. The quantum circuit for the first link $ \alpha=1.0 $ is presented in Fig.(\ref{fig:FIG-3}), while similar procedure is used for the rest of the links. First multi controlled $ X $ gate is used to identify the sub-state that relates to the source term for the link $ \alpha=1.0 $ using the second ancilla register $ a_2 $, while the next segment, which consist of five single controlled $ X $ gates (CNOT), and a multi-controlled $ X $ gate, is used to shift this state to a 'location' on which a Hadamard test can be applied. Following each of the 5 sub-states is 'aligned' with the corresponding sub-state containing $ g_{\alpha}^{eq} $, a Hadamard gate $ H $ is applied, which consequently result in quantum state
\begin{equation}
	{\begin{array}{rl}\ket{\phi_{1}}&\displaystyle=\frac{\ket{0}_{a2} \ket{0}_{a1}}{\sqrt{2}}\left( \sum_{k=0}^{5(N\times M)-1} a_{k,k} \lambda_{\alpha,k}/\| \lambda \| \ket{k}_q+\sum_{k=5(N\times M)-1}^{10(N\times M)-1} ({b_{k,k} \lambda_{\alpha,k}+c_{k,k}^*})/\| \lambda \| \ket{k}_q\right)\\
			\\
			&+\displaystyle\frac{\ket{0}_{a2}\ket{0}_{a1}}{\sqrt{2}}\left(\cdots \right)+\ket{10\lambda}_{a2a1q}+\ket{11\lambda}_{a2a1q}.
	\end{array}}
	\label{21}
\end{equation}
In contrast to Eq.(\ref{20}), prefactor $ 1/\sqrt{2} $ is inserted, while the second sum in the above equation is modified by introducing term $ c_{kk}^*=c_{ii} \lambda_{\alpha,i} $, where $ i=10(N\times M)-1,\ldots,15(N\times M)-1 $. The states of lesser interest are denoted as $ \ket{10\lambda}_{a2a1q}$ and $\ket{11\lambda}_{a2a1q} $. As a final result, a new term proportional to Eq.(\ref{9}) with $ \varepsilon=1.0 $ and $ \Delta t=1.0 $ is obtained.       
\begin{figure}
	\centering
	\begin{quantikz}
		\lstick{$q_{n-i}$} & \qwbundle [alternate]{} \slice{$ \ket{\phi_{0}} $} &  \gate[6, disable auto height][1.0cm]{B} \qwbundle [alternate]{}  & \qwbundle [alternate]{} & \qwbundle [alternate]{} \slice{$ \ket{\phi_{1}^*} $} & \qwbundle [alternate]{} & \qwbundle [alternate]{} & \qwbundle [alternate]{} & \qwbundle [alternate]{} & \qwbundle [alternate]{}  & \qwbundle [alternate]{} & \qwbundle [alternate]{}  & \qwbundle [alternate]{}  & \qwbundle [alternate]{} & \cdots & \qwbundle [alternate]{} \slice{$ \ket{\phi_{1}} $} & \qwbundle [alternate]{}     \\ 
		\lstick{$q_{n-4}$} & \qw &  \qw  & \qw & \qw & \qw & \octrl{1} & \qw & \qw & \qw & \qw & \targ{} & \ctrl{1} & \qw & \cdots & \qw & \qw  \\
		\lstick{$q_{n-3}$} & \qw &  \qw  & \qw & \qw & \qw & \ctrl{1} & \qw & \qw & \qw & \targ{} & \qw & \octrl{1} & \qw & \cdots & \qw & \qw  \\
		\lstick{$q_{n-2}$} & \qw &  \qw  & \qw & \qw & \qw & \octrl{1} & \qw & \qw & \targ{} & \qw & \qw & \ctrl{1} & \qw & \cdots & \qw & \qw  \\
		\lstick{$q_{n-1}$} & \qw &  \qw  & \qw & \qw & \qw & \ctrl{1} & \qw & \targ{} & \qw & \qw & \qw & \octrl{1} & \qw & \cdots & \qw & \qw  \\
		\lstick{$q_{n}$} & \qw &  \qw  & \targX{} & \qw & \qw & \octrl{1} & \qw & \qw & \qw & \qw & \qw & \octrl{1} & \qw & \cdots & \qw & \qw  \\ [0.5cm]
		\lstick{$a_1$} & \qw &  \gate{H}  & \swap{-1} & \gate{H}  & \qw & \octrl{1} & \targ{} & \qw & \qw & \qw & \qw & \ctrl{1} & \qw & \cdots &  \gate{H} & \qw \\
		\lstick{$a_2$} & \qw & \qw & \qw & \qw & \qw & \targ{} & \ctrl{-1} & \ctrl{-3} & \ctrl{-4} & \ctrl{-5} & \ctrl{-6} & \targ{} & \qw & \cdots & \qw & \qw 
	\end{quantikz}
	\fcaption{Quantum circuit for implementing the collision step.}
	\label{fig:FIG-3}
\end{figure}            
\begin{figure}
	\centering
	\begin{quantikz}[align equals at=1, column sep=0.4cm]
		& \gate{R}  \qwbundle [alternate]{} & \qwbundle [alternate]{}
	\end{quantikz}=\begin{quantikz} \lstick{$q_0$} & \ctrl{3} & \cdots & \ctrl{2} & \targ{} & \qw & \ctrl{3} & \qw \\
		\lstick{$q_1$}  & \ctrl{2} & \cdots & \ctrl{2} & \qw & \targ{} & \targ{} & \qw \\
		\lstick{$q_2$}  & \ctrl{2} & \cdots & \targ{} & \qw & \qw &  \qw & \qw \\
		\wave&&&&&&&\\ 
		\lstick{$q_{n-2}$}  & \ctrl{1} & \cdots & \qw & \qw & \qw &  \qw &   \qw & \\
		\lstick{$q_{n-1}$}  & \targ{} & \cdots & \qw & \qw & \qw &  \qw &   \qw & \\
		\lstick{$q_{n}$}  & \octrl{-2} & \cdots & \octrl{-3} & \octrl{-6} & \octrl{-5} & \octrl{-3} &  \qw &
	\end{quantikz}
	\fcaption{Quantum circuit for implementing the right shift.}
	\label{fig:FIG-4}
\end{figure}

\begin{figure}
	\centering
	\begin{quantikz}[align equals at=1, column sep=0.4cm]
		& \gate{L}  \qwbundle [alternate]{} & \qwbundle [alternate]{}
	\end{quantikz}=\begin{quantikz}                
		\lstick{$q_0$}      & \targ{} & \ctrl{3}  & \qw        & \ctrl{2} & \qw      & \cdots & \qw        & \ctrl{3}      &  \qw    & \qw \\
		\lstick{$q_1$}      & \qw      & \targ{}   & \qw        & \ctrl{2} & \qw      & \cdots & \qw        & \ctrl{2}     &  \qw    & \qw \\
		\lstick{$q_2$}      & \qw      & \qw       & \targ{}    & \targ{}  & \targ{}  & \cdots & \qw        & \ctrl{1}      &  \qw      & \qw \\
		\wave&&&&&&&&&&\\ 
		\lstick{$q_{n-2}$}  & \qw      & \qw       & \qw        & \qw      & \qw      & \cdots & \targ{}    & \targ{}  &  \targ{}  & \qw \\
		\lstick{$q_{n-1}$}  & \qw      & \qw       & \qw        & \qw      & \qw      & \cdots & \qw        & \qw      &  \qw      & \qw  \\
		\lstick{$q_{n}$}    & \ctrl{-6}& \ctrl{-4} & \ctrl{-4}  & \ctrl{-4}& \ctrl{-4}& \cdots & \ctrl{-2} & \ctrl{-3} & \ctrl{-2} &  \qw &
	\end{quantikz}
	\fcaption{Quantum circuit for implementing the left shift.}
	\label{fig:FIG-5}
\end{figure}        
\begin{figure}
	\centering
	\begin{quantikz}
		\lstick{$\log_2M$}  \slice{$ \ket{\phi_{1}} $} &  \qwbundle [alternate]{} \gategroup[wires=9, steps=4, style={dashed,	rounded corners, inner sep=0.2pt}]{vorticity $\omega$}  & \qwbundle [alternate]{} & \gate[wires=1]{L_{\alpha=3}} \qwbundle [alternate]{} & \gate[wires=1]{R_{\alpha=4}} \qwbundle [alternate]{} & \qwbundle [alternate]{} \gategroup[wires=9, steps=4, style={dashed,	rounded corners, inner sep=0.2pt}]{stream $ \psi $} & \qwbundle [alternate]{} & \gate[wires=1]{L_{\alpha=3}} \qwbundle [alternate]{} & \gate[wires=1]{R_{\alpha=4}} \qwbundle [alternate]{} \slice{$ \ket{\phi_{2}} $}  & \qwbundle [alternate]{} \\ 
		\lstick{$\log_2M$}  &  \gate[wires=1]{R_{\alpha=1}} \qwbundle [alternate]{}  & \gate[wires=1]{L_{\alpha=2}} \qwbundle [alternate]{} & \qwbundle [alternate]{} & \qwbundle [alternate]{} & \gate[wires=1]{R_{\alpha=1}} \qwbundle [alternate]{} & \gate[wires=1]{L_{\alpha=2}} \qwbundle [alternate]{} & \qwbundle [alternate]{} & \qwbundle [alternate]{}  & \qwbundle [alternate]{}   \\ 
		\lstick{$q_{n-4}$}  &  \octrl{-1}  & \ctrl{-1} & \ctrl{-2} & \ctrl{-2} & \ctrl{-1} & \octrl{-1} & \octrl{-2} & \ctrl{-2} & \qw    \\
		\lstick{$q_{n-3}$}  &  \octrl{-1}  & \octrl{-1} & \ctrl{-1} & \octrl{-1} & \octrl{-1} & \ctrl{-1} & \octrl{-1} & \ctrl{-1} & \qw   \\
		\lstick{$q_{n-2}$}  &  \octrl{-1}  & \octrl{-1} & \octrl{-1}& \octrl{-1} & \ctrl{-1} & \ctrl{-1} & \octrl{-1} & \ctrl{-1} & \qw    \\
		\lstick{$q_{n-1}$}  &  \octrl{-1}  & \octrl{-1}& \octrl{-1} & \octrl{-1} & \octrl{-1} & \octrl{-1} & \ctrl{-1} & \octrl{-1} & \qw   \\
		\lstick{$q_{n}$}  &  \octrl{-1}  & \octrl{-1} & \octrl{-1} & \octrl{-1} & \octrl{-1} & \octrl{-1} & \octrl{-1} & \octrl{-1} & \qw   \\ [0.5cm]
		\lstick{$a_1$}  &  \octrl{-1}  & \octrl{-1} & \octrl{-1}  & \octrl{-1} & \octrl{-1} & \octrl{-1} & \octrl{-1} & \octrl{-1} & \qw  \\
		\lstick{$a_2$}  & \octrl{-1} & \octrl{-1} & \octrl{-1} & \octrl{-1} & \octrl{-1} & \octrl{-1} & \octrl{-1} & \octrl{-1} & \qw 
	\end{quantikz}
	\fcaption{Quantum circuit for implementing the propagation step.}
	\label{fig:FIG-6}
\end{figure}                  

\emph{Propagation} - The propagation step implies shifting the previously calculated distribution functions $ f_{\alpha} $ and $ g_{\alpha} $ along four $ \alpha $ links with velocities $ e_{\alpha} $ defined by Eq.(\ref{5}). In other words, distribution functions that lie on horizontal links $ \alpha=1 $ and $ \alpha=2 $ are, according to Fig.(\ref{fig:FIG-1}), moved one lattice step in right and left direction, respectively, while in case of vertical links $ \alpha=3 $ and $ \alpha=4 $ this shift is done one lattice step up and down, respectively. The distribution functions for the rest particle $ \alpha=0 $ is kept stationary. In terms of quantum computing (QC) logic this spatial-temporal displacement of the distribution functions $ f_{\alpha} $ and $ g_{\alpha} $ can be achieved by applying unitary operation based on the quantum walk procedure \cite{Childs4}, where the quantum circuit for the right and left shifts are shown in Fig.(\ref{fig:FIG-4}) and Fig.(\ref{fig:FIG-5}), while the corresponding circuit for the entire propagation step is depicted in Fig.(\ref{fig:FIG-6}). Following the propagation step, a new state $ \phi_{2} $ is created 

\begin{equation}
	{\begin{array}{rl}\ket{\phi_{2}}&\displaystyle=\frac{\ket{0}_{a2} \ket{0}_{a1}}{\sqrt{2}}P_{\alpha}\left( \sum_{k=0}^{5(N\times M)-1} a_{k,k} \lambda_{\alpha,k}/\| \lambda \| \ket{k}_q+\sum_{k=5(N\times M)-1}^{10(N\times M)-1} ({b_{k,k} \lambda_{\alpha,k}+c_{k,k}^*})/\| \lambda \| \ket{k}_q\right)\\
			\\
			&+\displaystyle\frac{\ket{0}_{a2}\ket{0}_{a1}}{\sqrt{2}}\left(\cdots \right) +\ket{10\lambda}_{a2a1q}+\ket{11\lambda}_{a2a1q}.
	\end{array}}
	\label{22}
\end{equation}
where in case of links $ \alpha=1 $ and $ \alpha=4 $, operator $ P_{\alpha} $ is replaced by operator $ R $, while for tlinks $ \alpha=2 $ and $ \alpha=3 $, operator $ L $ is accordingly used.   

\emph{Macros} - In order to calculate macroscopic values according to Eq.(\ref{11}), point-wise addition of the distribution function $ f_{\alpha} $ in case of vorticity $ \omega $, and distribution function $ g_{\alpha} $ for the stream-function $ \psi $ is required. To perform this operation quantumly, a quantum circuit composed of controlled $ H $, $ SWAP $, and $ X $ gate is applied, where the $ X $ and $ SWAP $ gates are used primarily for the state preparation purposes, while an addition procedure is conducted by applying the  Hadamard operation $ H $. The corresponding quantum circuit for the calculation of the vorticity $ \omega $ is given in Fig.({\ref{fig:FIG-7}}), while the circuit for the stream-function $ \psi $ is depicted in Fig.(\ref{8}). At this point macroscopic variables of vorticity $ \omega $ and stream-function $ \psi $ are encoded into the quantum state 

\begin{equation}
	{\begin{array}{rl}\ket{\phi_{3}}&\displaystyle=\frac{\ket{0}_{a2} \ket{0}_{a1}}{4}\left( \sum_{k=0}^{(N\times M)-1} \omega_k^*/\| \lambda \| \ket{k}_q+\sum_{k=5(N\times M)-1}^{6(N\times M)-1} \psi_k^*/\| \lambda \| \ket{k}_q\right)\\
			\\
			&\displaystyle+\frac{\ket{0}_{a2} \ket{0}_{a1}}{2}\left( \sum_{k=15(N\times M)-1}^{16(N\times M)-1} \omega_{b(k)}^*/\| \lambda \| \ket{k}_q\right)+\ket{00\lambda}_{a2a1q} +\ket{10\lambda}_{a2a1q}+\ket{11\lambda}_{a2a1q}.
	\end{array}}
	\label{23}
\end{equation}
where term $ \omega^*/\| \lambda \| $ denotes the normalized amplitude proportional to the vorticity $ \omega $, $ \psi^*/\| \lambda \| $ is the amplitude proportional to the stream-function $ \psi $, while $ \omega_b^*/\| \lambda \| $ encodes the amplitudes of the boundary condition according to Eq. (\ref{14}). The rest of the state, denoted as $ \ket{00\lambda}_{a2a1q} +\ket{10\lambda}_{a2a1q}+\ket{11\lambda}_{a2a1q} $, does not contain any information relevant for the final outcome and it is primarily used for intermediate calculation purposes. 

\begin{figure}
	\centering
	\begin{quantikz}
		\lstick{$q_{n-i}$}  \slice{$ \ket{\phi_{2}} $} &  \qwbundle [alternate]{}  & \qwbundle [alternate]{} & \qwbundle [alternate]{} & \qwbundle [alternate]{} & \qwbundle [alternate]{} & \qwbundle [alternate]{} & \qwbundle [alternate]{} & \qwbundle [alternate]{}    & \qwbundle [alternate]{} & \slice{$\ket{\phi_{3}} $} \qwbundle [alternate]{} & \qwbundle [alternate]{}  \\  
		\lstick{$q_{n-4}$}  & \gate{H} & \gate{H} & \targX{} & \ctrl{1} & \ctrl{1} & \gate{H} & \targ{} & \ctrl{1}& \gate{H} & \octrl{1} & \qw   \\
		\lstick{$q_{n-3}$}  & \octrl{-1} & \ctrl{-1} & \octrl{-1} & \ctrl{1} & \gate{H} & \octrl{-1} & \ctrl{-1} & \ctrl{1} & \ctrl{-1} & \gate{H} & \qw  \\
		\lstick{$q_{n-2}$}  & \octrl{-1} & \octrl{-1} & \swap{-1} & \octrl{1} & \octrl{-1} & \octrl{-1} & \octrl{-1} & \octrl{1} & \octrl{-1} & \octrl{-1} & \qw  \\
		\lstick{$q_{n-1}$}  & \octrl{-1} & \octrl{-1} & \octrl{-1} & \octrl{1} & \octrl{-1} & \octrl{-1} & \octrl{-1} & \octrl{1} & \octrl{-1} & \octrl{-1} & \qw  \\
		\lstick{$q_{n}$}  & \octrl{-1} & \octrl{-1} & \octrl{-1} & \octrl{1} & \octrl{-1} & \octrl{-1} & \octrl{-1} & \octrl{1} & \octrl{-1} & \octrl{-1} & \qw   \\ [0.5cm]
		\lstick{$a_1$}  & \octrl{-1} & \octrl{-1} & \octrl{-1} & \octrl{1}& \octrl{-1} & \octrl{-1} & \octrl{-1} & \octrl{1} & \octrl{-1} & \octrl{-1} & \qw  \\
		\lstick{$a_2$}  & \qw & \qw & \qw & \targ{} & \qw & \qw & \ctrl{-1} & \targ{} & \qw & \qw & \qw
	\end{quantikz}
	\fcaption{Quantum circuit for calculating the vorticity $ \omega $.}
	\label{fig:FIG-7}
\end{figure}

\begin{figure}
	\centering
	\begin{quantikz}
		\lstick{$q_{n-i}$}  \slice{$ \ket{\phi_{2}} $} &  \qwbundle [alternate]{}  & \qwbundle [alternate]{} & \qwbundle [alternate]{} & \qwbundle [alternate]{} & \qwbundle [alternate]{} & \qwbundle [alternate]{} & \qwbundle [alternate]{} & \qwbundle [alternate]{}    & \qwbundle [alternate]{} & \qwbundle [alternate]{}     & \qwbundle [alternate]{} & \qwbundle [alternate]{} & \qwbundle [alternate]{} &  \qwbundle [alternate]{} & \slice{$\ket{\phi_{3}} $}  \qwbundle [alternate]{} & \qwbundle [alternate]{} \\  
		\lstick{$q_{n-4}$}  & \ctrl{1}   & \gate{H}   & \ctrl{1}   & \gate{H}   & \targ{}    & \ctrl{1}   & \ctrl{1}   & \gate{H}    & \octrl{1}  & \qw         & \qw       & \qw       & \targ{}   & \ctrl{1} & \ctrl{1} & \qw \\
		\lstick{$q_{n-3}$}  & \gate{H}   & \octrl{-1} & \ctrl{1}   & \ctrl{-1}  & \ctrl{-1}  & \gate{H}   & \octrl{1}  & \octrl{-1}  & \octrl{1}  & \qw         & \qw       & \targ{}   & \qw       & \ctrl{1} & \gate{H} & \qw \\
		\lstick{$q_{n-2}$}  & \ctrl{-1}  & \octrl{-1} & \ctrl{1}   & \ctrl{-1}  & \ctrl{-1}  & \ctrl{-1}  & \octrl{1}  & \octrl{-1}  & \octrl{1}  & \qw         & \targ{}   & \qw       & \qw       & \ctrl{1} & \ctrl{-1} & \qw\\
		\lstick{$q_{n-1}$}  & \octrl{-1} & \ctrl{-1}  & \octrl{1}  & \octrl{-1} & \octrl{-1} & \octrl{-1} & \ctrl{1}   & \ctrl{-1}   & \ctrl{1}   & \targ{}     & \qw       & \qw       & \qw       & \octrl{1} & \octrl{-1} & \qw \\
		\lstick{$q_{n}$}    & \octrl{-1} & \octrl{-1} & \octrl{1}  & \octrl{-1} & \octrl{-1} & \octrl{-1} & \octrl{1}  & \octrl{-1}  & \octrl{1}  & \qw         & \qw       & \qw       & \qw       & \octrl{1} & \octrl{-1} & \qw\\ [0.5cm]
		\lstick{$a_1$}      & \octrl{-1} & \octrl{-1} & \octrl{1}  & \octrl{-1} & \octrl{-1} & \octrl{-1} & \octrl{1}  & \octrl{-1}  & \octrl{1}  & \qw         & \qw       & \qw       & \qw       & \octrl{1} & \octrl{-1} & \qw \\
		\lstick{$a_2$}      & \qw        & \qw        & \targ{}    & \qw        & \qw        & \qw        & \targ{}    & \qw         & \targ{}    & \ctrl{-3}   & \ctrl{-4} & \ctrl{-5} & \ctrl{-6} & \targ{} & \qw & \qw
	\end{quantikz}
	\fcaption{Quantum circuit for calculating the stream-function $ \psi $.}
	\label{fig:FIG-8}
\end{figure}                                

\emph{Boundary conditions} - In the last step, implementation of the boundary conditions for both vorticity $ \omega $ and stream-function $  \psi  $ is conducted. In case of the vorticity, formulation defined by Eq.(\ref{15})  is used. Combining this equations with Eq.(\ref{11}), it becomes apparent that the macroscopic values for the vorticity on solid boundaries are reduced to much simpler form described by Eq.(\ref{14}). Practically, this is achieved by replacing vorticity $ \omega $ on the boundary lines with the corresponding stream function $  \psi  $ belonging to the first line inside the computational domain, with the optional addition of the velocity (this is the information that is actually encoded in term $ \omega_b^*/\| \lambda \| $). In case of the stream-function $ \psi $, zero values are assigned to the boundary lines. In order to manage this transformation quantumly, assigning zero to $ \omega $ and $ \psi $ along the boundary lines is conducted in the first step, while the implementation of Eq.(\ref{14}), which includes point-wise addition of the previously assigned zero values to $ \omega $, and the corresponding $ \psi $ and $ U $, are done in the second step. The agreement of the prefactors according to Eq.(\ref{23}) in the first step is followed by the procedure of zero-value implementation by applying the diagonal matrix using the Qiskit's diagonal subroutine \cite{Qiskit}. To equalize prefactors in Eq.(\ref{23}), two multi-controlled $ H $ gates are used, while the controlled form of the Hadamard test is applied in the case of the diagonal matrix implementation. Two multi-controlled $ X $ gates  are used in between for shifting the amplitudes related to the stream-function from position $ k=5(N\times M)-1 $ to $ k=2(N\times M)-1 $. The quantum circuit coresponding to the boundary condition is given in Fig.(\ref{fig:FIG-9}). Operator $ Cb $ is constructed from the diagonal matrix having entries that corresponds to zero values along boundary lines, and 1.0 everywhere else. Since this matrix is not unitary, the procedure used previously in the collision step is applied to the construction of the corresponding unitary matrices, which are then implemented as the upper and lower half of operator $ Cb $. After applying this segment of the quantum circuit onto state $ \ket{\psi_3} $, the amplitudes corresponding to vorticity $ \omega $ and stream-function $ \psi $ are set to zero along the boundary lines. Shifting the state defining the second part of the vorticity boundary condition ($ \omega_b^*/\| \lambda \| $), followed by the corresponding point-wise addition is manged in the second step. For the shift part, two multi-controlled $ X $ gates and one multi-controlled $ SWAP $ gate are used, while single $ H $ gate operated on the $ a_2 $ register is applied in case of the addition procedure. Finally, a quantum state proportional to vorticity $ \omega $ and stream-function $ \psi $  is obtained at the end of one time step

\begin{equation}
	{\begin{array}{rl}\ket{\phi_{4}}&\displaystyle=\frac{\ket{0}_{a2} \ket{0}_{a1}}{4\sqrt{2}}\left( \sum_{k=0}^{(N\times M)-1} \omega_k^{**}/\| \lambda \| \ket{k}_q+\sum_{k=(N\times M)-1}^{2(N\times M)-1} \psi_k^{**}/\| \lambda \| \ket{k}_q\right)\\
			\\
			&\displaystyle+\ket{00\lambda}_{a2a1q} +\ket{10\lambda}_{a2a1q}+\ket{11\lambda}_{a2a1q},
	\end{array}}
	\label{24}
\end{equation}
where $ \omega^{**} $ and $ \psi^{**} $ denote the amplitudes proportional to the vorticity and stream-function, respectively. The final values of macroscopic variables $ \omega $ and $ \psi $ are, according to Eq.(\ref{24}), obtained by re-normalization of the post-selected $ \ket{0}_{a2} \ket{0}_{a1} $ state by factor $ 4\| \lambda \|\sqrt{2} $.

\subsection{Post-processing}
To calculate the velocity components $ u $ and $ v $, defined by Eq.(\ref{3}), post-processing by using a classical computer is performed in this work. The main reason for using this approach here is not influenced by the complexity of the corresponding equation, which, on contrary, has a very simple structure, but by the technical limitation in terms of computational efficiency of the statevector simulator. To quantumly simulate Eq.(\ref{3}) a simple matrix having entries corresponding to the second-order discretization of the stream-function in $ x $ and $ y $ direction is initially created. Since this matrix is not unitary, a linear combination of the unitaries approach \cite{Low} can be applied. However, due to the collapse of the wave function during the measurements, two additional copies of quantum states, which contain the stream-function variable $ \psi $, is required. To provide these two copies of the stream-function, two approaches can be considered. According to the first approach the proposed quantum algorithm can be simultaneously executed on two additional quantum registers, where in each of these two register the corresponding matrix is applied following the boundary condition step. In general, these two additional registers do not require inclusion of two registers each having $ n $ qubits, but two quantum states created by inserting two additional qubits are sufficient. However, adding two more sets of quantum gates which operate according to the previously proposed algorithm significantly increases number of gates and consequently decreases the efficiency of the overall algorithm. Therefore, the second approach, in which the final state described by Eq.(\ref{24}) can be further transformed by using two additional qubits and the Hadamard operator, can be applied more efficiently. Splitting the current state into sets of quantum states are performed by using the $ H $ operator in the first step. Following the required number of states, proportional to state Eq.(\ref{24}), are being created, in the second step inclusion of the matrix of coefficients corresponding to the discretization of  Eq.(\ref{23}) by applying the linear combination of unitaries approach is conducted. As a consequence of using this method, statevectors proportional to the velocity components $ u $ and $ v $ are obtained by applying just two additional qubits and a minimal set of quantum gates. Following the velocity components are calculated, entries of the encoding diagonal matrix can be prepared for the calculation of the next time step. Similar procedure was addressed by Steijl and Barakos \cite{Steijl}.

To start with the next time step calculation, derivation of the quantum state vector, encoding the macroscopic variables from the previous time step in form of Eq.(\ref{18}), is required. In order to prepare such state, the process of tomography is performed, after which the encoding using  previously proposed procedure is conducted. Along with the state vector encoding, preparation of the diagonal matrix having the entries corresponding to Eq.(\ref{6}) and Eq.(\ref{7}) is also executed. This 'classical' matrix preparation is imposed by the non-linear term in Eq.(\ref{6}), i.e. the product of two dependent variables, velocity and vorticity. Both of these procedures are performed on a classical machine, making the proposed quantum algorithm to certain extent a quantum-classical hybrid. Hence, in order to achieve full speed-up of the quantum algorithm for the Navier-Stokes equations, the process of tomography and non-linearity needs to be properly addressed.                                     

\subsection{Complexity}   
To analyze the complexity of the proposed quantum algorithm, an approach of dividing the analysis into LBM steps used in an earlier work \cite{Budinski} is adopted in this paper. In general, the encoding section and the two diagonal operators, one for collision and one for the boundary condition step, are the most demanding procedures in terms of the input dimension-quantum gates relation. According to Shende et al. \cite{Shende}, disentangling the qubits in the preparation of quantum states requires $ 2^n-1 $ steps and 1 diagonal operator, which, as stated by Theorem 7 in \cite{Shende}, can be constructed from $ 2\times 4^n-\left(2n+3\right)\times 2^n+2n $ CNOT gates. Furthermore, in case of the diagonal operator used for the collision and boundary condition step, the total CNOT count, according to the same author is $ 2\times 2^n $. The rest of the circuit corresponding to the collision and boundary step does not depend on the size of the input, hence it relates to the number of qubits as $\mathcal{O}(1)$. For the propagation step, the number of multi-controlled $ X $ gates scales with the number of qubits for each operator $ R $ and $ L $ as $ \mathcal{O} (n) $, meaning that each additional qubit requires one additional multi-controlled $ X $ gate per operator with the number of controls increased by one (Fig.(4)). Since there are four $ R $ and four $ L $ operators in the proposed algorithm, the total number of additional multi-controlled $ X $ gates per one added qubit for the propagation step is eight. The control operations per operators $ R $ and $ L $ are not influenced by the size of the system. The final step includes the calculation of the macroscopic variables which is not affected by the size of the input domain. To increase the number of computational points by adding more qubits this particular segment of the algorithm does not require any additional gates, hence it is also scaled with the number of qubits as $ \mathcal{O} (1) $. It is obvious from this complexity analysis that the encoding step should be further investigated in order to reduce number of the CNOT gates and hence lower the input size-number of gate dependence. One of the approaches that can be considered in this algorithm is to eliminate the need for any kind of state re-normalization performed at the end of each computational step, in which the preparation step will be performed just once at the beginning of simulation. This could completely eliminate the preparation step and hence significantly increase the efficiency of the algorithm. On the other side, performing the quantum state tomography after each time step severely influences the efficiency of the algorithm. Due to the collapse of the quantum state after each measurement, the algorithm needs to prepare $ 2^{2n} $ input states  for each time step, i.e. each time step needs to be repeated $ 2^{2n} $ times before the algorithm can move on the next step. This confirms previously stated fact that the presented algorithm solves single time step, while the consecutive transfer from one time step to another is the major obstacle that yet needs to be resolved. However, due to its complexity, the implementation of this methodology will be objective of future work.         

\section{Validation}
To validate the proposed algorithm a cavity flow problem is simulated using the \textit{‘statevector simulator’} backend as a part of the Qiskit \cite{Qiskit} platform, and compared with the results obtained by the FORTRAN code \cite{Xavier}. For this particular problem a square domain with $ 16 \times 16 $ lattice configuration is used, where the time and lattice step in $ x $ and $ y $ direction are set as $ \Delta t=1 $ and $ \Delta x=\Delta y=1.0 $, respectively. For the relaxation parameter value of $ \varepsilon=1 $ is adopted, while the velocity of moving wall is set to $ U=1.0 $, all in lattice units. Zero vorticity and stream-function is used as initial condition, while steady solution is achieved after 500 computational steps. Comparison of the results obtained by the proposed algorithm and the FORTRAN code are in form of vorticity contours and velocity magnitude as given in Fig.(\ref{10}). Exact agreement between the compared values is obtained.           

\begin{figure}
	\centering
	\begin{quantikz}
		\lstick{$q_{n-i}$}  \slice{$ \ket{\phi_{3}} $} &  \qwbundle [alternate]{}  & \qwbundle [alternate]{} & \qwbundle [alternate]{} & \qwbundle [alternate]{} & \qwbundle [alternate]{} & \gate[8, disable auto height][1.0cm]{Cb} \qwbundle [alternate]{}  & \qwbundle [alternate]{} & \qwbundle [alternate]{}    & \qwbundle [alternate]{} & \qwbundle [alternate]{}     & \slice{$ \ket{\phi_{4}} $} \qwbundle [alternate]{} & \qwbundle [alternate]{} \\  
		\lstick{$q_{n-4}$}  & \gate{H}   & \qw   & \qw   & \ctrl{1}   & \qw     & \qw   & \qw   & \qw   & \qw  & \octrl{1}         & \qw    &   \qw \\
		\lstick{$q_{n-3}$}  & \ctrl{-1}   & \gate{H} & \qw   & \octrl{1}  & \qw   & \qw   & \qw  & \targ{}  & \qw & \octrl{1}        & \qw    &   \qw \\
		\lstick{$q_{n-2}$}  & \ctrl{-1}  & \ctrl{-1} & \octrl{1}   & \targ{}  & \octrl{1}  & \qw  & \octrl{1}  & \octrl{-1}  & \targX{}  & \octrl{1}          & \qw &  \qw\\
		\lstick{$q_{n-1}$}  & \ctrl{-1} & \ctrl{-1}  & \octrl{1}  & \octrl{-1} & \octrl{1} & \qw & \octrl{1}   & \octrl{-1}   & \ctrl{2}   & \targ{}     & \qw   &    \qw \\
		\lstick{$q_{n}$}    & \octrl{-1} & \octrl{-1} & \targ{}  & \octrl{-1} & \octrl{1} & \qw & \octrl{1}  & \octrl{-1}  & \octrl{-2}  & \octrl{-1}         & \qw   &    \qw\\ [0.5cm]
		\lstick{$a_1$}      & \octrl{-1} & \octrl{-1} & \octrl{-1}  & \octrl{-1} & \octrl{1} & \qw & \octrl{1}  & \octrl{-1}  & \octrl{1}  & \octrl{-1}        & \qw       & \qw \\
		\lstick{$a_2$}      & \qw        & \qw        & \ctrl{-1}   & \octrl{-1}  & \gate{H}        & \qw        & \gate{H}    & \ctrl{-1}        & \targX{}    & \ctrl{-1}   & \gate{H} & \qw
	\end{quantikz}
	\fcaption{Quantum circuit for implementing the boundary conditions.}
	\label{fig:FIG-9}
\end{figure}              

\begin{figure}

	\centering
	\subfloat[][]{
		\begin{tikzpicture}
			\begin{axis}[title={vorticity $ \psi $}, xmin=0, xmax=15, ymin=0, ymax=15, view={0}{90}, legend pos=south west,  legend image post style={
					sharp plot,draw=\pgfkeysvalueof{/pgfplots/contour/draw color},
				}, legend entries={classic,quantum}]
				\addplot [contour prepared={levels={-0.7533, -0.6636, -0.5739, -0.4842, -0.3945, -0.3048, -0.2151, -0.1254},draw color=red,labels=false,}, ultra thick,] file {FVORT.dat};
				\addplot [contour prepared={levels={-0.7533, -0.6636, -0.5739, -0.4842, -0.3945, -0.3048, -0.2151, -0.1254},draw color=lime,contour label style={/pgf/number format/fixed,/pgf/number format/precision=3,every node/.append style={text=black}}, label distance=200pt}, semithick,] file {QVORT.dat};
				
			\end{axis}
		\end{tikzpicture}
		\label{fig:subim1}}
	\quad
	\subfloat[][]{
		\begin{tikzpicture}
			\begin{axis}[title={velocity magnitude $ v_{mag} $}, xmin=0, xmax=15, ymin=0, ymax=15, view={0}{90}]
				\addplot [contour prepared={levels={0.0556, 0.1111, 0.1667, 0.2222, 0.2778, 0.3333, 0.3889, 0.4444},draw color=red,labels=false,}, ultra thick,] file {FVELM.dat};
				\addplot [contour prepared={levels={0.0556, 0.1111, 0.1667, 0.2222, 0.2778, 0.3333, 0.3889, 0.4444},draw color=lime,contour label style={/pgf/number format/fixed,/pgf/number format/precision=3,every node/.append style={text=black}}, label distance=150pt}, semithick,] file {QVELM.dat};
				
			\end{axis}
		\end{tikzpicture}
		\label{fig:subim2}}
	
	\fcaption{Comparison of the numerical results, obtained by the quantum algorithm and the classical  FORTRAN code, for cavity flow simulated by the D2Q5 LBM.}
	\label{fig:FIG-10}
\end{figure} 	    

\section{Conclusion} 
A novel quantum algorithm for solving the two-dimensional Navier-Stokes equations using the vorticity-stream function formulation and the lattice Boltzmann method as a numerical procedure is presented in this work. Both of the equations are modeled using the D2Q5 lattice configuration, where for each computational step a corresponding quantum circuit is constructed and presented in detail. To test and validate the proposed quantum algorithm practically the IBM's open source quantum platform Qiskit is used to simulate the 2D cavity flow induced by moving the top wall. As a result of this simulation a quantum state proportional to the solution of the vorticity-stream function equations is obtained and correspondingly compared with he classically calculated values. Excellent agreement between the compared values is achieved confirming that the quantum device is capable of  solving efficiently complex problems of fluid flows described by the NSE. Since the state preparation procedure in the encoding step is marked as the most demanding in terms of efficiency (same fact is observed and reported in quantum machine learning experiments), further research will be oriented towards the elimination of the state encoding steps between two consecutive time steps, which is expected to improve significantly the efficiency of the overall algorithm.              

\nonumsection{References}

\end{document}